\documentclass{PoS}

\def\beq{\begin{equation}}
\def\eeq{\end{equation}}
\def\bea{\begin{eqnarray}}
\def\eea{\end{eqnarray}}
\def\beqn{\begin{eqnarray}} \def\eeqn{\end{eqnarray}}
\def\beeq{\begin{eqnarray}}
\def\eeeq{\end{eqnarray}}

\def\nn{\nonumber}

\def\td#1{\tilde{\delta}\left(#1\right)}

\def\ii{\imath 0}

\def\uv{{\rm UV}}

\def\r{{\rm R}}

\title{
\vspace*{-2.0cm}
On the universal structure of Higgs amplitudes mediated by heavy particles}

\ShortTitle{On the universal structure of Higgs amplitudes mediated by heavy particles}

\author{\speaker{Germ\'an F. R. Sborlini}$^{\ a,b}$, F\'elix Driencourt-Mangin$^b$ and Germ\'an Rodrigo$^b$\\\\
        $^a$Dipartimento di Fisica, Universit\`a di Milano and INFN Sezione di Milano,
I-20133 Milan, Italy.\\\
        $^b$Instituto de F\'{\i}sica Corpuscular, Universitat de Val\`{e}ncia -- 
Consejo Superior de Investigaciones Cient\'{\i}ficas, Parc Cient\'{\i}fic, E-46980 Paterna, Valencia, Spain.\\\

        E-mail: \email{german.sborlini@unimi.it,felix.dm@ific.uv.es,german.rodrigo@csic.es}}

\abstract{We describe the calculation of the one-loop corrections to $H \to \gamma \gamma$ and $g g \to H$ within the four-dimensional unsubtraction/loop-tree duality (FDU/LTD) approach. The fact that these corrections are both IR and UV finite is not enough to perform the computation without a proper regularization of the intermediate steps. We show how the FDU/LTD method unambiguously leads to the proper result with a pure four-dimensional representation of the loop amplitude. Moreover, this method allows us to obtain very compact expressions which share the same functional structure independently of the particles circulating the loop. Besides this, asymptotic expansions for the low and high mass regime naturally arise as a consequence of the smoothness of the integrands obtained to describe the amplitudes.}

\FullConference{EPS-HEP 2017, European Physical Society conference on High Energy Physics\\
		5-12 July 2017\\
		Venice, Italy}

\begin{document}

\section{Introduction}
\label{sec:intro}
The one-loop amplitudes for Higgs to massless gauge bosons are finite because there is no a tree-level
contribution in the Standard Model. However, the expressions involved are individually ill-defined
and the four-dimensional limit has to be considered carefully. In other terms, a finite result does not
guarantee that we can leave aside the regularization framework. In fact, a non-consistent four-dimensional
implementation would produce misleading results. Since the Higgs amplitudes exhibit this pathological behaviour, we used them as a case of study for our method: the four-dimensional unsubtraction (FDU) framework.

We center the discussion in two processes: Higgs production through gluon fusion and Higgs decay into a photon pair. Whilst in the first case only massive coloured fermions can be included in the loop, the diphoton decay also involves W bosons. We show in Ref. \cite{Driencourt-Mangin:2017gop} that the structure of all these contributions is universal, i.e. independent of the nature of the particle inside the loop. However, the specific UV behaviour of the different contributions is slightly different: this leads to potential discrepancies when naively considering the computation in four dimensions \cite{Weinzierl:2014iaa}. This problem can be easily solved within the FDU framework because we achieve a fully local regularization.

\subsection{Four-dimensional Unsubtraction (FDU) and Loop-Tree Duality (LTD)}
\label{ssec:intro1}
The FDU framework is based on three key ideas: the loop-tree duality (LTD) theorem, a momentum mapping between the virtual and real kinematics, and the introduction of local counter-terms. The LTD establishes that loop amplitudes can be expressed as the sum of dual amplitudes, which are built starting from tree-level like objects and replacing the loop measure with a phase-space one. Explicitly, the master formula is given by
\beq
\int_\ell \, \prod_{i=1}^N \, G_F(q_i) \equiv - \sum_{i=1}^N \, \int_\ell \tilde{\delta}(q_i) \, \prod_{j\neq i} G_D(q_i;q_j) \, ,
\eeq
for any scalar one-loop Feynman integral, with $G_F(q_i)=1/(q_i^2-m_i^2+ \imath 0)$ and the dual propagators $G_D(q_i;q_j)=1/(q_j^2 -m_j^2-\imath 0 \, \eta \cdot k_{ji})$ \cite{Catani:2008xa}. Notice that the prescription inside the dual propagators involves an arbitrary space-like vector $\eta$ and depends on both the cut line and the momentum flow associated to the original propagator. Moreover, the theorem is valid beyond one-loop and, also, when multiple powers of the propagators are present \cite{Bierenbaum:2010cy,Bierenbaum:2012th}. The last statement is crucial for the definition of local renormalization counter-terms, as we will briefly mention in the following section. 

Aside the purely theoretical developments, the LTD theorem has been succesfully applied to compute numerically many tensorial multi-leg loop amplitudes \cite{Buchta:2015wna}. The implementation is very efficient since the expressions to be integrated do not contain any Gram determinant (which are usually responsible of many numerical instabilities) and there are local cancellation among dual contributions that render the whole integrand smooth.

On the other hand, the method relies on the possibility of regularizing the expressions at integrand level
and in four space-time dimensions. To deal with UV singularities, we need integrand-level formulae
for the renormalization factors, while the real-emission is directly used as a local counter-term
in the IR region through the application of a proper momentum mapping \cite{Hernandez-Pinto:2015ysa,Sborlini:2016gbr,Sborlini:2016hat}. This mapping is motivated by the universal factorization of IR singularities in the scattering amplitudes \cite{Buchta:2014dfa,PROCFDU}.

\section{Results and discussion}
\label{ssec:results}
In order to apply the LTD/FDU to the Higgs amplitudes, we introduce a tensor basis and then define two projectors, i.e. 
\beq
P_1^{\mu \nu} = \frac{1}{d-2} \left(\eta^{\mu \nu}-\frac{2 p_1^{\nu}p_2^{\mu}}{s_{12}}-(d-1)\frac{2 p_1^{\mu}p_2^{\nu}}{s_{12}}\right)\, , \ \ \ \ P_2^{\mu \nu} = \frac{2 p_1^{\mu} p_2^{\nu}}{s_{12}} \, ,
\eeq
which allow us to extract the relevant scalar coefficients. The momenta of the photons/gluons is $p_1$ and $p_2$ with $s_{12} = (p_1+p_2)^2$. Whilst the first projector is associated to the only non-trivial contribution to the physical amplitude, the second one vanishes, because of Ward's identities, after integration. So, we define ${\cal A}_{i}^{(1,f)}=P_i^{\mu \nu} {\cal A}_{\mu \nu}^{(1,f)}$ as the one-loop projected scalar amplitudes.

Then, we apply the LTD decomposition into dual contributions. In this case, there are four different propagators, so we obtain four dual terms associated to the internal momenta $q_1=\ell+p_1$, $q_2=\ell+p_{12}$, $q_3=\ell$ and $q_4=\ell+p_2$. Since the internal on-shell momenta can be related by unifying the coordinate system, we add the four contributions and obtain
\beqn
{\cal A}_1^{(1,f)} &=& \, g_f \, \int_\ell \, \td{\ell} \bigg[
\bigg( \frac{\ell_{0}^{(+)}}{q_{1,0}^{(+)}} + \frac{\ell_{0}^{(+)}}{q_{4,0}^{(+)}} +  \frac{2 \, (2\ell\cdot p_{12})^2}{s_{12}^2-(2\ell\cdot p_{12}-\ii)^2} \bigg) 
\bigg( \frac{s_{12}\, M_f^2}{(2\, \ell \cdot p_1)(2\, \ell \cdot p_2)}\, c_1^{(f)} \nn \\ &+&  c_2^{(f)} \bigg) 
+ \frac{2 \, s_{12}^2}{s_{12}^2-(2\ell\cdot p_{12}-\ii)^2} \, c_3^{(f)} 
\label{eq:uni1} 
\bigg]~, \\
{\cal A}_2^{(1,f)} &=& \, g_f  \, \frac{c_3^{(f)}}{2} \, \int_\ell \, \td{\ell} \, 
\bigg( \frac{\ell_{0}^{(+)}}{q_{1,0}^{(+)}} + \frac{\ell_{0}^{(+)}}{q_{4,0}^{(+)}} - 2 \bigg)~,
\label{eq:uni2}
\eeqn
with $f=\{\phi,t,W\}$ denoting the flavour of the particle circulating the loop. It is important to appreciate that the functional structure of the amplitudes is independent of the particle inside the loop; all the information is contained in the coefficients $c_i^{(f)}$ \cite{Driencourt-Mangin:2017gop}. We can further simplify ${\cal A}_1^{(1,f)}$ by exploiting the fact that ${\cal A}_2^{(1,f)}$ vanishes after integration. Thus, we get
\beqn
{\cal A}_1^{(1,f)} &=& \, g_f \, s_{12}\, \int_\ell \td{\ell} \bigg[
\bigg( \frac{\ell_{0}^{(+)}}{q_{1,0}^{(+)}} + \frac{\ell_{0}^{(+)}}{q_{4,0}^{(+)}} + \frac{2 \, (2\ell\cdot p_{12})^2}{s_{12}^2-(2\ell\cdot p_{12}-\ii)^2} \bigg)
\frac{M_f^2}{(2\, \ell \cdot p_1)(2\, \ell \cdot p_2)}\, c_1^{(f)}  \nn \\
&+& \frac{2 \, s_{12}}{s_{12}^2-(2\ell\cdot p_{12}-\ii)^2}\, c_{23}^{(f)} \bigg]~,
\label{eq:uniFIN}
\eeqn
which depends only on $c_1^{(f)}$ and $c_{23}^{(f)}=c_2^{(f)}+c_3^{(f)}$. Notice that the first line in Eq. (\ref{eq:uniFIN}) is integrable in four-dimensions, but the second one is UV divergent. However, this UV singularity is multiplied by $c_{23}^{(f)}$, which is proportional to $(d-4)$; a non-rigorous implementation of the four-dimensional limit might lead to discrepancies in the finite parts of the amplitude.

In order to remove the ambiguities related to the four-dimensional limit, we will introduce a local renormalization counter-term. Even if the full Higgs amplitude is UV finite, Eq. (\ref{eq:uniFIN}) shows that the integrand is locally singular in the high-energy region. The local UV counter-term is obtained by expanding the \emph{dual integrand around the UV propagator}, as explained in Refs. \cite{Driencourt-Mangin:2017gop,Sborlini:2016gbr}. Explicitly, we write 
\beqn
\left. {\cal A}^{(1,f)}_{1,\r} \right|_{d=4} &=& \left( {\cal A}^{(1,f)}_1 - {\cal A}^{(1,f)}_{1,\uv} \right)_ {d=4}\, ,
\label{eq:Amp1}
\\ {\cal A}^{(1,f)}_{1,\uv} &=& -   g_f  \, s_{12} \, \int_{\vec{\ell}}  \frac{1}{4 (q_{\uv,0}^{(+)})^3}
\Bigg( 1 + \frac{1}{(q_{\uv,0}^{(+)})^2} \frac{3\, \mu_\uv^2}{d-4} \Bigg)\, c_{23}^{(f)} \, ,
\label{eq:Amp2}
\eeqn
where the sub-leading term in the UV counter-term is defined to mimic the conventional DREG result (i.e. ${\cal A}^{(1,f)}_{1,\uv}=0$).

\subsection{Asymptotic expansions}
\label{ssec:results1}
Another important feature of the LTD approach is related with the possibility of performing expansions directly at integrand level. This is due to two properties of the expressions that we obtain. On one side, the integration measure is defined in an Euclidean space (namely, the spatial components of the loop-momentum) which allows to define a norm for the integration variable. On the other, we are always dealing with locally regularized expressions, that are integrable. In consequence, we guarantee the commutativity between asymptotic expansions and integration.   

In Ref. \cite{Driencourt-Mangin:2017gop}, we used the Higgs amplitudes at one-loop as a benchmark example to show the feasibility of the asymptotic expansions in the LTD/FDU framework. To have an insight about how the method works, let's consider the limit $M^2_f \gg s_{12}$ and use the expansion
\beq
\frac{\ell^{(+)}_0}{q_{1,0}^{(+)}} = \left(1-4\left(-\frac{2 \vec{\ell}\cdot \vec{p}_1+\vec{p}_1^2}{(2\ell^{(+)}_0)^2}\right)\right)^{-1/2}= \sum_{n=0} \frac{\Gamma(2n+1)}{\Gamma^2(n+1)} \left(-\frac{2 \vec{\ell}\cdot \vec{p}_1+\vec{p}_1^2}{(2\ell^{(+)}_0)^2}\right)^n\, ,
\eeq
with $q_{1,0}^{(+)}=\sqrt{(\vec{\ell}+\vec{p}_1)^2+M_f^2}$ and $\ell_0^{(+)}=\sqrt{\vec{\ell}^2+M_f^2}$. These expressions are valid because $\ell^{(+)}_0 > M_f$ for any value of the loop three-momentum. 

Using these formulae in each term of Eq. (\ref{eq:Amp1}), we manage to obtain the low and high mass limits at integrand level in Ref. \cite{Driencourt-Mangin:2017gop}. Of course, after integration, we recovered the well-known results available in the literature \cite{Aglietti:2006tp}. In particular, we succeeded in recovering the logarithmic factors that appear when $M_f \to 0$, without any additional handling of the expansions.

\section{Conclusions and outlook}
\label{sec:conclu}
We applied the LTD/FDU framework to tackle the computation of one-loop scattering amplitudes for Higgs bosons. The compact expressions obtained for the dual contributions allowed to achieve a universal integrand-level representation of the complete virtual amplitude. The information related with the nature of the virtual particles inside the loop is encoded in constant coefficients but the functional form is shared. This universality might be present even at higher-orders due to the properties of the LTD representation.
Also, we managed to remove the scheme ambiguities by implementing a local renormalization inspired by the FDU method \cite{Gnendiger:2017pys}. All the previously known results were recovered within a pure four-dimensional framework. Another important result of our work is the possibility of an straightforward implementation of asymptotic expansions at integrand-level. This is due to the fact that the dual integration is performed in an Euclidean space, namely the loop three-momentum space.

Finally, we expect further simplifications of higher-order computations and asymptotic expansions by exploiting the four-dimensional nature of the LTD/FDU approach.

\section*{Acknowledgments}
This work is partially supported by the Spanish Government, EU ERDF funds (grants FPA2014-53631-C2-1-P and SEV-2014-0398), by Generalitat Valenciana (GRISOLIA/2015/035) and Fondazione Cariplo under the grant number 2015-0761.



\begin{thebibliography}{99}

\bibitem{Driencourt-Mangin:2017gop}
  F.~Driencourt-Mangin, G.~Rodrigo and G.~F.~R.~Sborlini,
  arXiv:1702.07581 [hep-ph].

\bibitem{Weinzierl:2014iaa}
  S.~Weinzierl,
  Mod.\ Phys.\ Lett.\ A {\bf 29} (2014) no.15,  1430015.
  
\bibitem{Catani:2008xa}
  S.~Catani, T.~Gleisberg, F.~Krauss, G.~Rodrigo and J.~C.~Winter,
  JHEP {\bf 0809} (2008) 065.
 
\bibitem{Bierenbaum:2010cy}
  I.~Bierenbaum, S.~Catani, P.~Draggiotis and G.~Rodrigo,
  JHEP {\bf 1010} (2010) 073.
	
\bibitem{Bierenbaum:2012th}
  I.~Bierenbaum, S.~Buchta, P.~Draggiotis, I.~Malamos and G.~Rodrigo,
  JHEP {\bf 1303} (2013) 025.
  
\bibitem{Buchta:2015wna}
  S.~Buchta, G.~Chachamis, P.~Draggiotis and G.~Rodrigo,
  Eur.\ Phys.\ J.\ C {\bf 77} (2017) no.5,  274.
  

\bibitem{Hernandez-Pinto:2015ysa}
  R.~J.~Hern\' andez-Pinto, G.~F.~R.~Sborlini and G.~Rodrigo,
  JHEP {\bf 1602} (2016) 044.

\bibitem{Sborlini:2016gbr}
  G.~F.~R.~Sborlini, F.~Driencourt-Mangin, R.~J.~Hern\'andez-Pinto and G.~Rodrigo,
  JHEP {\bf 1608} (2016) 160.
 
\bibitem{Sborlini:2016hat}
  G.~F.~R.~Sborlini, F.~Driencourt-Mangin and G.~Rodrigo,
  JHEP {\bf 1610} (2016) 162.

\bibitem{Buchta:2014dfa}
  S.~Buchta, G.~Chachamis, P.~Draggiotis, I.~Malamos and G.~Rodrigo,
  JHEP {\bf 1411} (2014) 014.
  
\bibitem{PROCFDU}
  G. Sborlini, F. Driencourt-Mangin, R. Hern\'andez-Pinto and G. Rodrigo,
  PoS {\bf EPS-HEP2017} (2017) 547. TIF-UNIMI-2017-7.
  
\bibitem{Aglietti:2006tp}
  U.~Aglietti, R.~Bonciani, G.~Degrassi and A.~Vicini,
  JHEP {\bf 0701} (2007) 021.
  
\bibitem{Gnendiger:2017pys}
  C.~Gnendiger {\it et al.},
  Eur.\ Phys.\ J.\ C {\bf 77} (2017) no.7,  471.
 
\end{thebibliography}
\end{document}